\documentclass[12pt]{article}
\usepackage[a4paper]{geometry}
\usepackage{amsfonts,amsmath,amssymb}
\usepackage{mathrsfs,frcursive}
\usepackage{color}
\usepackage{theorem,cite}
\usepackage{hyperref}

\newtheorem{thm}{Theorem}[section]
\newtheorem{prop}[thm]{Proposition}
\newtheorem{lemma}[thm]{Lemma}

\newtheorem{rmk}{Remark}[section]

\newcommand{\bea}{\begin{eqnarray}}
\newcommand{\eea}{\end{eqnarray}}
\newcommand{\beali}{\begin{align}}
\newcommand{\eeali}{\end{align}}
\newcommand{\beano}{\begin{eqnarray*}}
\newcommand{\eeano}{\end{eqnarray*}}
\newcommand{\beq}{\begin{equation}}
\newcommand{\eeq}{\end{equation}}

\def\fs{{\mathfrak s}}

\def\cP{{\cal P}}  \def\cQ{{\cal Q}}

\newcommand{\II}{{\mathbb I}}

\newcommand{\jcur}{{j}}

\newcommand{\wt}[1]{\widetilde{#1}}

\newcommand{\half}{{\textstyle{\frac{1}{2}}}}

\newcommand{\proof}{\textbf{Proof:}~}
\newcommand{\qed}{{\hfill \rule{5pt}{5pt}}\\}

\def\diag{\mathop{\rm diag}\nolimits}

\def\sgn{\mathop{\rm sgn}\nolimits}

\def\tr{\mathop{\rm tr}\nolimits}

\newcommand{\doi}[1]{\href{https://doi.org/#1}{doi:#1}}
\newcommand\atopn[2]{\genfrac{}{}{0pt}{}{#1}{#2}}

\numberwithin{equation}{section}

\begin{document}
\pagestyle{empty}

\null
\vspace{20pt}

\parskip=6pt

\begin{center}
\begin{LARGE}
\textbf{Integrable quadratic structures\\[1.ex]
 in peakon models}
\end{LARGE}

\vspace{50pt}

\begin{large}
{J.~Avan${}^{a}$, L.~Frappat${}^b$, E.~Ragoucy${}^b$ \footnote[1]{emails: jean.avan@cyu.fr, luc.frappat@lapth.cnrs.fr, eric.ragoucy@lapth.cnrs.fr}}
\end{large}

\vspace{15mm}

${}^a$ \textit{Laboratoire de Physique Th\'eorique et Mod\'elisation,} \\
\textit{CY Cergy Paris Universit\'e, CNRS, F-95302 Cergy-Pontoise, France} \\

\vspace{5mm}

${}^b$ \textit{Laboratoire d'Annecy-le-Vieux de Physique Th{\'e}orique LAPTh,} \\ 
\textit{USMB, CNRS, F-74000 Annecy}

\end{center}

\vspace{4mm}

\begin{abstract}
We propose realizations of the Poisson structures for the Lax representations of three integrable $n$-body peakon equations, Camassa--Holm, Degasperis--Procesi and Novikov. The Poisson structures derived from the integrability structures of the continuous equations yield quadratic forms for the $r$-matrix representation, with the Toda molecule classical $r$-matrix playing a prominent role. 

We look for a linear form for the $r$-matrix representation. Aside from the Camassa--Holm case, where the structure is already known, the two other cases do not allow such a presentation, with the noticeable exception of the Novikov model at $n=2$.

Generalized Hamiltonians obtained from the canonical Sklyanin trace formula for quadratic structures are derived in the three cases.
\end{abstract}


\newpage

\baselineskip=16.5pt
\parskip=6pt
\parindent=0pt
\pagestyle{plain}
\setcounter{page}{1}

\section{Introduction}

Peakon solutions to non-linear two-dimensional $(x,t)$ integrable fluid equations have been shown to exhibit themselves integrable dynamics in several interesting cases. They take the generic form
\begin{equation}
\label{eq:peakonphi}
\varphi(x,t) = \sum_{i=1}^n p_i(t)\, e^{\vert x-q_i(t) \vert }
\end{equation}
and their dynamics for $(p_i,q_i)$ is deduced from a reduction of the 1+1 fluid equations for $\varphi(x,t)$.

Integrability of peakons entails the existence of a Poisson structure for $(p_i,q_i)$ deduced from the original Poisson structure of the fluid fields, including $\varphi(x,t)$; the existence of a Hamiltonian $h(p_i,q_i)$ and Poisson-commuting higher Hamiltonians, a priori deduced by reduction of the continuous Hamiltonians to peakon solutions for the integrable 1+1 dynamical system. In a number of cases, the dynamics is expressed in terms of a Lax equation:
\begin{equation}
\label{eq:lax}
\dot L = [ L,M ]
\end{equation}
where $L,M$ are $(p,q)$-dependent matrices and \eqref{eq:lax} contains all equations for $p_i(t)$, $q_i(t)$ obtained from plugging \eqref{eq:peakonphi} into the 1+1 integrable equation.

The Lax matrix naturally yields candidate conserved Hamiltonians $h^{(k)} = \tr(L^k)$. Poisson-commutation of $h^{k)}$ is equivalent \cite{BV} to the existence of an $r$-matrix formulation of the Lax matrix Poisson brackets:
\begin{equation}
\label{eq:PBL1L2}
\{ L_1 , L_2 \} = \sum \{ L_{ij} , L_{jk} \}\, e_{ij} \otimes e_{kl} = [ r_{12} , L_1 ] - [ r_{21} , L_2 ].
\end{equation}
The $r$-matrix itself may depend on the dynamical variables \cite{STS}. 
A simple example of such dynamical $r$-matrix is given by the reformulation of the well-known ``quadratic'' $r$-matrix structure, extensively studied \cite{Skl1982,Skl1983,MF}. We recall the form of this quadratic structure:
\begin{equation}
\label{eq:PBFM}
\{ L_1 , L_2 \} = a_{12} L_1 L_2 - L_1 L_2 d_{12} + L_1 b_{12} L_2 - L_2 c_{12} L_1
\end{equation}
where $a_{12} = -a_{21}$, $d_{12} = -d_{21}$, $b_{12} = c_{21}$ to ensure antisymmetry of the Poisson bracket. 
When the \textsl{regularity condition}\footnote{The name 'regularity' will be motivated in section \ref{sect:twist}.}
\begin{equation}\label{eq:trace0}
a_{12}-c_{12}=d_{12}-b_{12}
\end{equation}
 is fulfilled, \eqref{eq:PBFM} is indeed identified with \eqref{eq:PBL1L2} by setting $r_{12} = \half(a_{12}L_2+L_2a_{12})-L_2c_{12}$. Hence when the regularity condition 
\eqref{eq:trace0} is fulfilled, the quantities $\tr L^k$ mutually Poisson-commute, ensuring the integrability of the peakon models. 
The case $a=d$, $b=c=0$ was first characterized by E.~Sklyanin \cite{Skl1979}; $a=d$, $b=c$ yields the so-called classical reflection algebra \cite{Skl1982,Skl1983}.

We consider here the three integrable peakon equations discussed in e.g. \cite{AR2016} for which the key features of Poisson structure, integrability and Lax matrix, have been established: 

1. The Camassa--Holm equation \cite{CH,CHH}. Poisson structure for peakons is given in \cite{dGHH}, Lax formulation in \cite{RB} although the Poisson structure here is not the one in \cite{dGHH}. We shall comment and relate the two structures in Section 2.

2. The Degasperis--Procesi equation \cite{DGP,dGPro3}. Poisson structure for peakons is given also in \cite{dGHH}. Lax formulation is given in \cite{dGPro3}, also commented in \cite{HH}.

3. The Novikov equation \cite{Novi1}. Poisson structure for peakons is given in \cite{HW}. Lax formulation is given in \cite{HLS}.

Note that a fourth peakon-bearing integrable equation was identified (so-called modified Camassa--Holm equation \cite{Fuchs,Fokas}), but peakon integrability properties are obstructed by the higher non-linearity of the modified Camassa--Holm equation, precluding the consistent reduction of Poisson brackets and Hamiltonians to peakon variables \cite{AK}.

We will establish in these three cases the existence of a quadratic $r$-matrix structure \eqref{eq:PBFM}.

We will show that the four parametrizing matrices $a$, $b$, $c$, $d$ are equal or closely connected to the Toda $A_n$ $r$-matrix \cite{OT}. This close connection can be understood in the Camassa--Holm and Novikov cases, via an identification between the Lax matrix of Camassa--Holm and the well-known Toda molecule Lax matrix \cite{FO}. In addition, the construction of the Novikov Lax matrix as $L^{\text{Nov}} = TL^{\text{CH}}$, where $T = \sum_{i,j}^n \big(1+\sgn(q_i-q_j)\big) e_{ij}$, relates the two $r$-matrices by a twist structure. The occurence of $a_{12}$ in the Camassa--Holm context however also requires an understanding of the Camassa--Holm peakons Poisson bracket in \cite{dGHH} as a second Poisson bracket in the sense of Magri \cite{Magri,OR}, where the first Poisson bracket is the canonical structure $\{q_i,p_j\}=\delta_{ij}$, to be detailed in Section 2.

Each following section is now devoted to one particular model, resp. Camassa--Holm (Section 2), Degasperis--Procesi (Section 3), and Novikov (Section 4). We conclude with some comments and open questions.

\section{Camassa--Holm peakons\label{sect2}}

The Camassa--Holm shallow-water equation reads \cite{CH,CHH}
\begin{equation}
u_t - u_{xxt} + 3 uu_x = 2u_x u_{xx} + uu_{xxx}
\end{equation}
The $n$-peakon solutions take the form
\begin{equation}
u(x,t) = \sum_{i=1}^n p_i(t)\, e^{-|x-q_i(t)|}
\end{equation}
yielding a dynamical system for $p_i,q_i$:
\begin{equation}
\dot{q}_i = \sum_{j=1}^n p_j \, e^{-|q_i-q_j|} \,, \qquad \dot{p}_i = \sum_{j=1}^n p_ip_j \,\fs_{ij}\, e^{-|q_i-q_j|} \,.
\end{equation}
This discrete dynamical system is described by a Hamiltonian
\begin{equation}
H = \frac12\,\sum_{i,j=1}^n {p_ip_j} \,e^{ -| q_i-q_j| }
\end{equation}
 such that
\begin{equation}
\dot{f}=\{ f\,,\,H \} \,,
\end{equation}
with the canonical Poisson structure:
\begin{equation}\label{eq:PB-can}
\{ p_i,p_j \} = \{ q_i,q_j \} = 0 \,, \qquad \{ q_i,p_j \} = \delta_{ij} \,.
\end{equation}
The same dynamics is in fact also triggered \cite{dGHH} by the reduced Camassa--Holm Hamiltonian:
\begin{equation}
H = \sum_{i} p_i
\end{equation}
with the reduced Camassa--Holm Poisson structure (which is dynamical and ``non-local''):
\begin{equation}\label{eq:PBCH} 
\begin{aligned}
\{ {p}_i,{p}_j \} &= \fs_{ij}\,{p}_i {p}_j e^{-|{q}_i-{q}_j| } \,, \\
\{ {q}_i,{p}_j \} &= {p}_j e^{-|{q}_i-{q}_j|} \,, \\
\{ {q}_i,{q}_j \} &= \fs_{ij} \big( 1-e^{-|{q}_i-{q}_j| }\big) \,, 
\end{aligned}
\end{equation}
where $\fs_{ij}=\sgn({q}_i-{q}_j)$.
It is also encoded in the Lax formulation \cite{CF,CH}
\begin{equation}
\frac{dL}{dt} = [L,M]
\end{equation}
with
\begin{equation}
\label{eq:LCH}
L = \sum_{i,j=1}^n L_{ij} e_{ij} \,, \qquad L_{ij} = \sqrt{p_ip_j} \,e^{-\half | q_i-q_j | } \,.
\end{equation}

\subsection{The linear Poisson structure}
We summarize here the results obtained in \cite{RB}.
The Poisson structure \eqref{eq:PB-can} endows the Lax matrix \eqref{eq:LCH} with a linear $r$-matrix structure
\begin{equation}\label{eq:rlin-CH}
\{L_1\,,\, L_2\} = [r_{12}\,,\, L_1] -[r_{21}\,,\, L_2]\,,\quad\mbox{with}\quad r_{12}=a_{12}-b_{12}\,.
\end{equation}
In \eqref{eq:rlin-CH},  $a_{12}$ is the $A_{n-1}$ Toda $r$-matrix 
\begin{equation}\label{eq:Toda}
a_{12}=\frac14\,\sum_{i,j=1}^n \fs_{ij}\, e_{ij}\otimes e_{ji}=-a_{21} \qquad \text{and} \qquad b_{12}=-a_{12}^{t_2} \,,
\end{equation}
with by convention $\sgn(0) = 0$, and $e_{ij}$ is the $n\times n$ elementary matrix with 1 at position $(i,j)$ and 0 elsewhere. Connection of the Lax matrix with the Toda $r$-matrix structure was already pointed out in \cite{RB}. The $r$-matrix structure \eqref{eq:rlin-CH} is indeed identified with the same structure occuring in the so-called Toda lattice models \cite{OR}. 
One can add that
the $r$-matrix structure for the Toda lattice in \cite{RB} and the peakon dynamics in \eqref{eq:rlin-CH} is directly 
identified with the well-known $r$-matrix structure for Toda molecule models \cite{FO}. Indeed, both Toda lattice  and 
peakon Lax matrices endowed with the canonical 
Poisson structure \eqref{eq:PB-can} are representations of the abstract $A_{n-1}$ Toda molecule structure 
\begin{equation}
L = \sum_i x_i\,h_i +\sum_{\alpha\in\Delta_+} x_\alpha (e_{\alpha}+e_{-\alpha})\,,
\end{equation}
with
$\{x_\alpha\,,\,x_\beta\} = x_{\alpha+\beta}$ and $\{h_i\,,\,x_\alpha\} = \alpha(i)\,x_{\alpha}$.

In the highly degenerate case of Toda Lax matrix (where $x_\alpha=0$ for non-simple roots $\alpha$), it is directly checked 
that $a$ and $s$ yield the same contribution to \eqref{eq:rlin-CH}, implying that the Toda Lax matrix has an $r$-matrix structure 
parametrized by $a_{12}$ solely, as it is well-known \cite{OT}.

\subsection{The quadratic Poisson structure}

The new result which we shall elaborate on now is stated as:
\begin{prop}\label{prop:CHQ}
The Poisson structure \eqref{eq:PBCH} endows the Lax matrix \eqref{eq:LCH} with a quadratic $r$-matrix structure:
\begin{equation}
\label{eq:LLQuad}
\{ L_1 , L_2 \} = [ a_{12} , L_1 L_2 ] - L_2 b_{12} L_1 + L_1 b_{12} L_2,
\end{equation}
where $a_{12} $ and $b_{12}$ are given in \eqref{eq:Toda}.
\end{prop}
\proof Direct check by computing the Poisson bracket $\{ L_{ij},L_{kl}\}$ on the left hand side and right hand side. The antisymmetry of the Poisson structure, explicitly realized by \eqref{eq:LLQuad}, allows to eliminate ``mirror display'', i.e. $(ij,kl) \leftrightarrow (kl,ij)$.
The invariance of the Poisson structure \eqref{eq:LLQuad} under each operation $t_1$ and $t_2$ is due to the symmetry $L^t = L$ of \eqref{eq:LCH}, the identification of $b_{12} = -a_{12}^{t_2}$, and the antisymmetry $a_{12}^{t_1t_2} = -a_{12}$. It allows to eliminate transposed displays $(ij,kl) \leftrightarrow (ji,kl) \leftrightarrow (ji,lk) \leftrightarrow (ij,kl)$ and to check only a limited number of cases (indeed 13 cases).

Remark that the form \eqref{eq:LLQuad} ensures that the regularity condition \eqref{eq:trace0} is trivially obeyed.

 The Poisson structure \eqref{eq:PBCH}, identified as a second Poisson structure
 in the sense of Magri \cite{Magri},  yields the natural quadratization \eqref{eq:LLQuad}  of the $r$-matrix structure \eqref{eq:rlin-CH}.
 This fact is consistent with the fact that \eqref{eq:PBCH} is obtained by reduction to peakon variables of the second Poisson structure of {Camassa--Holm}, built in \cite{dGHH}, while \eqref{eq:PB-can} is obtained by reduction of the first Camassa--Holm Poisson structure.
 Reduction procedure (from fields to peakon variables) and recursion construction (\textit{\`a la} Magri, see \cite{OR}) are therefore compatible in this case, and the compatibility extends to the $r$-matrix structures of the reduced variables.
 Such a consistency at the $r$-matrix level is not an absolute rule. For instance, the first and second Poisson structures for the Calogero--Moser model yield $r$-matrix structures, ``linear'' \cite{ABT} and ``quadratic'' \cite{AR}, but with different $r$-matrices.

\subsection{The Yang--Baxter relations}
\paragraph{Quadratic structure.} As is known from general principles \cite{MF}, quadratic Poisson $r$-matrix structures obey consistency quadratic equations of Yang--Baxter type to ensure Jacobi identity of Poisson brackets. 
In the case of original Camassa--Holm pair $(a,b)$ in \eqref{eq:Toda}, the skew-symmetric element $a_{12}$ obeys the modified Yang--Baxter equation:
\begin{equation}
\label{eq:modYB}
[ a_{12},a_{13} ] + [ a_{12},a_{23} ] + [ a_{13},a_{23} ] = \frac1{16}\Big(\Omega_{123} - \Omega_{123}^{t_1t_2t_3}\Big) \,,
\ \text{ with }\ \Omega_{123}=\sum_{i,j,k=1}^n e_{ij}\otimes e_{jk}\otimes e_{ki} \,.
\end{equation}

 The symmetric element $b_{12}$ obeys an adjoint-modified Yang--Baxter equation directly obtained from transposing \eqref{eq:modYB} over space 3:
\begin{equation}
\label{eq:modYBadj}
[ a_{12},b_{13} ] + [ a_{12},b_{23} ] + [ b_{13},b_{23} ] = \frac1{16}\Big(- \Omega_{123}^{t_3} + \Omega_{123}^{t_1t_2}\Big) \,.
\end{equation}
Cancellation of a suitable combination of \eqref{eq:modYB} and \eqref{eq:modYBadj} with all permutations added, together with symmetry of the Camassa--Holm Lax matrix, allows to then check explicitly Jacobi identity for $L^{CH}$ and Poisson structure \eqref{eq:LLQuad}. \\

\paragraph{Linear structure.} The Jacobi identity for the linear Poisson structure \eqref{eq:rlin-CH} also follows from \eqref{eq:modYB} and \eqref{eq:modYBadj}. Associativity of the linear Poisson bracket is equivalent to the cyclic relation:

\begin{equation}
\label{eq:Jacobi-lin}
[[ r_{12},r_{13} ] + [ r_{12},r_{23} ] + [ r_{32},r_{13} ], L_1] + cyclic = 0\,,
\end{equation}
where \textit{cyclic} stands for sum over cyclic permutations of $(1,2,3)$.

The Yang--Baxter "kernel" $[ r_{12},r_{13} ] + [ r_{12},r_{23} ] + [ r_{32},r_{13} ]$ must now be evaluated. In many models, it is known to be equal to $0$ (classical Yang--Baxter equation) or to a combination of the cubic Casimir operators $\Omega_{123}$ and $\Omega_{123}^{t_1t_2t_3} $ 
(modified Yang--Baxter equation). If any of these two sufficient conditions holds, \eqref{eq:Jacobi-lin} is then trivial. However, in the Camassa--Holm case, the situation is more involved. Indeed  from  \eqref{eq:modYB} and \eqref{eq:modYBadj}, and denoting the Casimir term 
$C_{123} \equiv \Omega_{123} - \Omega_{123}^{t_1t_2t_3}$, one has:

\begin{equation}
\label{eq:Jacobi-lin2}
[ r_{12},r_{13} ] + [ r_{12},r_{23} ] + [ r_{32},r_{13} ] = C_{123} +  C_{123}^{t_3}  + C_{123}^{t_2} - C_{123}^{t_1}
\end{equation}

which is neither a cubic Casimir nor even cyclically symmetric. Realization of \eqref{eq:Jacobi-lin} indeed follows from explicit direct cancellation of the first (factorizing) Casimir term in \eqref{eq:Jacobi-lin2} under commutation with $L_1 + L_2 + L_3$ and cross-cancellation of the remaining 9 terms, using in addition the invariance of $L$ under transposition. We have here a textbook example of an $r$-matrix parametrizing a Poisson structure for a Lax matrix without obeying one of the canonical classical Yang--Baxter equations.

\section{Degasperis--Procesi peakons}\label{sect3}

This integrable shallow-water equation reads \cite{dGPro3}
\begin{equation}
u_t - u_{xxt} + 4 uu_x = 3 u_x u_{xx} + u u_{xxx}
\end{equation}
Note that, together with the Camassa--Holm equation, it is a particular case of the so-called $b$-equations:
\begin{equation}
u_t - u_{xxt} + (\beta+1) uu_x = \beta u_x u_{xx} + u u_{xxx}
\end{equation}
for which integrability properties are established for $\beta=2$ (Camassa--Holm) and $\beta=3$ (Degasperis--Procesi), by an asymptotic integrability approach \cite{dGPro3}. This approach fails at $\beta=4$. 

\subsection{The quadratic Poisson structure}
For $\beta=3$, $n$-peakon solutions are parametrized as
\begin{equation}
u(x,t) = \half \sum_{j=1}^n p_j(t)\, e^{-|x-q_j(t)| } \,,
\end{equation}
yielding a dynamical system:
\begin{equation}
\label{eq:dynGP}
\begin{split}
\dot{p}_j &= 2 \sum_{k=1}^n p_j p_k \,\fs_{jk}\, e^{ -|q_j-q_k|} \\
\dot{q}_j &= \sum_{k=1}^n p_k\, e^{-|q_j-q_k| } \,.
\end{split}
\end{equation}
Note the extra factor 2 in $\dot{p}_j$ compared with the Camassa--Holm equation. \\
The Lax matrix is now given by
\begin{equation}
\label{eq:LaxGP}
L_{ij} = \sqrt{p_ip_j} \, \big( T_{ij} - \fs_{ij}\,e^{-|q_i-q_j|} \big) \,,
\end{equation}
with 
\begin{equation}
\label{eq:defT}
T_{ij} = 1+\fs_{ij}\,, \quad i,j=1,\dots,n\,.
\end{equation}
The dynamical equations \eqref{eq:dynGP} derive from the Hamiltonian $H = \tr(L)$ and the Poisson structure obtained by reduction from the canonical Poisson structure of Degasperis--Procesi:
\begin{align}
\{ p_i,p_j \} &= 2{p}_i {p}_j \fs_{ij}\, e^{-|{q}_i-{q}_j|} \,, \nonumber \\
\label{eq:PBDGP} \{ q_i,p_j \} &= {p}_j e^{-|{q}_i-{q}_j|} \,, \\
\{ q_i,q_j \} &= \half  \fs_{ij}\, \big( 1-e^{-|{q}_i-{q}_j|}\big) \,. \nonumber
\end{align}
Again one notes the non-trivial normalization of the Poisson brackets in \eqref{eq:PBDGP} compared with \eqref{eq:PBCH} which will have a very significant effect on the $r$-matrix issues.
Let us note that the Hamiltonian associated to a time evolution $\dot{f}=\{ f\,,\,H \}$ consistent with the dynamics \eqref{eq:dynGP} is in fact the conserved quantity noted $P$ in \cite{dGPro3}. The Hamiltonian $H$ in \cite{dGPro3} is $\tr L^2$.

Let us now state the key result of this section.

\begin{prop}\label{prop:DGP}
The Poisson structure \eqref{eq:PBDGP} endows the Lax matrix $L$ given in \eqref{eq:LaxGP} with a quadratic $r$-matrix structure:
\begin{equation}
\label{eq:LLDGP}
\{ L_1 , L_2 \} = [ a'_{12} , L_1 L_2 ] - L_2 b'_{12} L_1 + L_1 b'_{12} L_2 \,,
\end{equation}
where (we remind our convention that $\sgn(0)=0$)  
\begin{align}
a'_{12} &= \frac12\sum_{i,j} \fs_{ij}\, e_{ij} \otimes e_{ji}=2\,a_{12}\, , \\
b'_{12} &= -\frac12\sum_{i,j} \fs_{ij}\, e_{ij} \otimes e_{ij} -\frac12\cQ_{12} \,,\quad \text{with}\quad \cQ_{12}= \sum_{i,j=1}^n e_{ij} \otimes e_{ij}.
\end{align}
\end{prop}

\proof: By direct check of the left hand side and right hand side of \eqref{eq:LLDGP}.
Since the Lax matrix $L$ is neither symmetric nor antisymmetric, many more cases of inequivalent index displays occur.  More precisely, 12 four-indices, 18 three-indices and 5 two-indices  must be checked.
\qed

The regularity condition \eqref{eq:trace0} is again trivially fulfilled.

\subsection{The Yang--Baxter equations}
The matrix $a'_{12}$ in the Degasperis--Procesi model is essentially the linear Toda $r$-matrix as in the Camassa--Holm model. On the contrary, the $b'$ component of the quadratic structure \eqref{eq:LLDGP} must differ from the $b$ component in the quadratic Camassa--Holm bracket \eqref{eq:LLQuad} by an extra term proportional to $\cQ_{12} =\cP_{12}^{t_1}=\cP_{12}^{t_2}=\cQ_{21}$, where 
$\cP_{12}= \sum_{i,j=1}^n e_{ij} \otimes e_{ji}$ is the permutation operator between space 1 and space 2.
For future use, we also note the property
\begin{equation}
\cP_{12}^2=\II_n\otimes\II_n \quad\text{and}\quad \cP_{12} \, M_1 \, M'_2\,\cP_{12} = M'_1 \, M_2 \,,
\end{equation}
\begin{equation}\label{eq:propQ}
M_1\,\cQ_{12} = M^t_2\,\cQ_{12} \quad\text{and}\quad \cQ_{12}\,M_1=\cQ_{12}\,M^t_2
\end{equation}
which holds for any $n\times n$ matrices $M$ and $M'$.
\begin{rmk}
Since the Lax matrix $L^{CH}$ of Camassa--Holm peakons is a symmetric $c$-number matrix, $L=L^t$, one checks that
\begin{equation}
\cP_{12}\, L_1\, L_2 = L_2\, L_1\, \cP_{12}  \quad\text{and}\quad L_1\, \cP_{12}^{t_1}\, L_2 = L_2\, \cP_{12}^{t_1}\, L_1\,.
\end{equation}
Hence the $(a',b')$ pair of $r$-matrices yielding the quadratic Poisson structure for Degasperis--Procesi peakons yields the quadratic Poisson structure for Camassa--Holm peakons with a pair $(a,b+\frac14\cQ)$, since the extra contribution $L_2 \,\cQ_{12}\, L_1 - L_1\, \cQ_{12}\, L_2$ cancels out. In this case, we will call this pair an alternative presentation for Camassa--Holm peakons.
\end{rmk}

The Degasperis--Procesi $(a',b')$ pair obeys a  set of classical Yang--Baxter equations which is simpler than the Camassa--Holm pair $(a,b)$. 
The skew-symmetric element $a_{12}$ still obeys a modified Yang--Baxter equation 
\begin{equation}
[ a'_{12},a'_{13} ] + [ a'_{12},a'_{23} ] + [ a'_{13},a'_{23} ] = \frac1{4}\Big(\Omega_{123} - \Omega_{123}^{t_1t_2t_3}\Big) \,,
\end{equation}
 but the symmetric element $b'$ obeys an adjoint-modified Yang--Baxter equation with zero right-hand-side:
\begin{equation}
\label{eq:YBb}
[ a'_{12},b'_{13} ] + [ a'_{12},b'_{23} ] + [ b'_{13},b'_{23} ] =0 \,.
\end{equation}
\begin{rmk}
A term proportional to $\cP_{12}$ can be added to $a'_{12}$, leading to a matrix $\wt a'_{12}=a'_{12}+\frac12\cP_{12}$. This term is optional, it does not change the Poisson brackets, nor the regularity condition. If added, it allows the relation $b'_{12}=-(\wt a'_{12})^{t_2}$, which already occurred for the Camassa--Holm model. 
However, such a term breaks the antisymmetry relation $a'_{21}=-a'_{12}$, which has deep consequences at the level of Yang--Baxter equations. 
Indeed, the form of the left-hand-side in \eqref{eq:modYB} heavily relies on this antisymmetry property of $a'_{12}$. In fact, if one computes
"naively" $[ \wt a'_{12},\wt a'_{13} ] + [ \wt a'_{12},\wt a'_{23} ] + [ \wt a'_{13},\wt a'_{23} ]$, one finds exactly zero and could be tempted to associate it to a Yang--Baxter equation with zero right-hand-side. Yet, the "genuine" Yang--Baxter equation, i.e. the relation ensuring the associativity of the Poisson brackets, plugs the $\Omega$-term back into the game, leading \textit{in fine} to again a  modified Yang--Baxter equation.
\end{rmk}

\subsection{Search for a linear Poisson structure}
Contrary to the Camassa--Holm peakon case, the canonical Poisson bracket \eqref{eq:PB-can} is not compatible with the soliton-derived Poisson bracket structure \eqref{eq:PBDGP}. Indeed, the linear pencil $\{\cdot\,,\,\cdot\}_{\text{can}} + \lambda \{\cdot\,,\,\cdot\}_{DP}$ (``can'' is for canonical and DP for Degasperis--Procesi) does not obey Jacobi identity due to extra non-cancelling contributions from the non-trivially \emph{scaled} brackets of $\{p,p\}$ and $\{q,q\}$ in \eqref{eq:PBDGP}. The statement is consistent with the fact, pointed out in \cite{dGHH}, that no second local Poisson structure exists in the Degasperis--Procesi case, contrary to the Camassa--Holm case (where it is denoted as $B_1$ in \cite{dGHH}).

Consistently with the absence of a second local Poisson structure for soliton Degasperis--Procesi equation yielding a ``linear'' $r$-matrix structure for the Lax matrix, one observes that the associated linear $r$-matrix structure  naively defined by $r_{12}=a'_{12}+b'_{12}$, does not yield consistent Poisson brackets for the variables in the Lax matrix. If one indeed sets
\begin{equation}
\{ L_1 , L_2 \} = [ a'_{12}+b'_{12} , L_1 ] - [ a'_{12}+b'_{12} , L_2 ] \,,
\end{equation}
the Poisson brackets for individual coordinates of $L$ are inconsistent, due to the antisymmetric part in \eqref{eq:LaxGP}, contrary to the Camassa--Holm case where $L_{ij}=L_{ji}$.

We also checked using software calculations that (at least for $n$ running from 2 to 7) that there is no
non-trivial linear combination $r'_{12}=x\,a'_{12}+y\,b'_{12}$ such that the relation 
$\{L_1\,,\,L_2\}=[r'_{12}\,,\,L_1]-[r'_{21}\,,\,L_2]$ with $L$ given in \eqref{eq:LaxGP}, yield a consistent Poisson structure for the $(p_i,q_j)$ variables. 

The peakon Lax matrix \eqref{eq:LaxGP} realizes therefore an interesting example of a non-dynamical quadratic $(a',b')$ Poisson structure 
where there is no associated linear $r$-matrix structure. The exact form, or even the existence, of such 
linear $r$-matrix structure for Degasperis--Procesi peakons remains an open question.

\section{Novikov peakons}
The Novikov shallow-wave equation reads
\begin{equation}
u_t - u_{xxt} + 4 u^2u_x = 3 uu_x u_{xx} + u^2 u_{xxx},
\end{equation}
showing now a cubic non-linearity instead of a quadratic one as in Camassa--Holm or Degasperis--Procesi. Originally proposed by Novikov \cite{MN} as an integrable partial differential equation, it was later shown \cite{HW} to have integrable peakons:
\begin{equation}
u(x,t) = \sum_{i=1}^n p_i(t)\, e^{-|x-q_i(t)|}.
\end{equation}

\subsection{The quadratic Poisson structure}

The complete integrability structure was established in \cite{HLS}. The dynamical system for $p_i,q_i$ reads
\begin{equation}
\label{eq:dynNov}
\begin{split}
\dot{p}_i &= p_i \sum_{j,k=1}^n \fs_{ij}\,p_j p_k \, e^{-|q_i-q_j|-|q_i-q_k|} \\
\dot{q}_i &= \sum_{j,k=1}^n p_j p_k\, e^{-|q_i-q_j|-|q_i-q_k| } \,,
\end{split}
\end{equation}
still with the notation $\fs_{ij}=\sgn(q_i-q_j)$.
They constitute a Hamiltonian system where the Poisson structure takes the following form:
\begin{equation}\label{eq:PBNov} 
\begin{aligned}
\{ p_i,p_j \} &= \fs_{ij}\,p_i p_j\,  e^{-2|q_i-q_j| } \,, \\
\{ q_i,p_j \} &= p_j \,e^{-2|q_i-q_j| } \,, \\
\{ q_i,q_j \} &= \fs_{ij}\, \big( 1-e^{-2|q_i-q_j| }\big) \,, 
\end{aligned}
\end{equation}
The conserved Hamiltonians are obtained as traces of a Lax matrix
\begin{equation}
L = TPEP
\end{equation}
where
\begin{equation}
T_{ij} = 1+\fs_{ij}, \qquad 
P_{ij} = p_i \delta_{ij}, \qquad
E_{ij} = e^{-|q_i-q_j| }.
\end{equation}
In other words, the time evolution \eqref{eq:dynNov} is described by the Hamilton equation $\dot f = \{f\,,\,H\}$, 
with $H=\frac12\tr L$ and the PB \eqref{eq:PBNov}.

Redefining now
\begin{equation}\label{eq:RenNov}
\bar{q}_j = 2q_j \qquad \text{and} \qquad \bar{p}_j = p_j^2,
\end{equation}
yield a Poisson structure
\begin{equation}\label{eq:PBNov2} 
\begin{aligned}
\{ \bar{p}_i,\bar{p}_j \} &= 4\fs_{ij}\,\bar{p}_i \bar{p}_j e^{ -|\bar{q}_i-\bar{q}_j|} \,, \\
\{ \bar{q}_i,\bar{p}_j \} &= 4\bar{p}_j e^{ -|\bar{q}_i-\bar{q}_j| } \,, \\
\{ \bar{q}_i,\bar{q}_j \} &= 4\fs_{ij}\, \big( 1-e^{ -|\bar{q}_i-\bar{q}_j|}\big) \,, 
\end{aligned}
\end{equation}
identical to the Camassa--Holm peakon structure \eqref{eq:PBCH} up to a factor 4. The Lax matrix now reads
\begin{equation}
\label{eq:LaxNov}
L_{ij} = \sum_{k=1}^n T_{ik} \sqrt{\bar{p}_k\,\bar{p}_j} \, e^{ -\half|\bar{q}_j-\bar{q}_k| }
\end{equation}
exactly identified with $TL^{CH}$.

Hence, the Novikov peakons are in fact described by a Lax matrix simply twisted from the Camassa--Holm Lax matrix ($L \to TL$) and an identical Poisson bracket, a fact seemingly overlooked in \cite{HLS}.
The $r$-matrix structure immediately follows, but several inequivalent structures are identified due to the gauge covariance pointed out in section \ref{sect3}:
\begin{prop}
The Poisson structure \eqref{eq:PBNov} endows the Lax matrix \eqref{eq:LaxNov} with a set of quadratic $r$-matrix structure
\begin{equation}\label{eq:L1L2Nov}
\{ L_1,L_2 \} = a''_{12} L_1L_2 - L_1L_2 d''_{12} + L_1b''_{12}L_2 - L_2c''_{12}L_1  \,,
\end{equation}
where
\begin{equation}
\label{eq:twistabcd}
\begin{aligned}
& a''_{12} = 4\, T_1\,T_2\, a_{12}\, T_1^{-1}\,T_2^{-1}, \quad && d''_{12} =  4\,a_{12}\,, \\
& b''_{12} = T_2 \, \big( -4a_{12}^{t_2}- \cQ_{12}\big)\, T_2^{-1}, \qquad &&c''_{12} = b''_{21}\,,
\end{aligned}
\end{equation}
and $a_{12}$ is given in \eqref{eq:Toda}.
\end{prop}
The proof follows trivially from section 2 and gauge invariance in section 3.

The regularity condition \eqref{eq:trace0} is fulfilled by this Poisson structure. Although less trivial than in the previous two cases this property will be proved in the next section.

\subsection{The Yang--Baxter equations}
The Yang--Baxter equations for \eqref{eq:twistabcd} follow immediately by suitable conjugations by $T$ of the Yang--Baxter equations for the alternative form of Degasperis--Procesi structures matrices. Precisely, from the redefinitions in \eqref{eq:twistabcd} the Yang--Baxter equations for $a''$, $b''$, $c''$ and $d''$ read
\begin{equation}
\label{eq:ybe-Novi}
\begin{aligned}
& [ a''_{12},a''_{13} ] + [ a''_{12},a''_{23} ] + [ a''_{13},a''_{23} ] = \Omega_{123} - \Omega_{123}^{t_1t_2t_3} \,,\\
& [ d''_{12},d''_{13} ] + [ d''_{12},d''_{23} ] + [ d''_{13},d''_{23} ] = \Omega_{123} - \Omega_{123}^{t_1t_2t_3} \,,\\
&[ a''_{12},b''_{13} ] + [ a''_{12},b''_{23} ] + [ b''_{13},b''_{23} ] =0 \,,\\
&[ d''_{12},c''_{13} ] + [ d''_{12},c''_{23} ] + [ c''_{13},c''_{23} ] =0 \,,
\end{aligned}
\end{equation}
where in writing the right-hand-side of the relation for $a''$, we have used the property that $\Omega_{123} - \Omega_{123}^{t_1t_2t_3}$ commutes with any product of the form $M_1\,M_2\,M_3$ for any matrix $M$ ($M=T$ for the present calculation).
Note that the adjoint Yang--Baxter equations for $b''$ and $c''$ remain with zero right-hand-side, despite the conjugations by $T$ depend on the matrices (e.g. $a''$ or $b''$) one considers.

\paragraph{Linear structure.}
As for the Degasperis--Procesi case, we looked for a linear combination $r''_{12}= x\,a''_{12}+y\,b''_{12}+z\,c''_{12}+t\,d''_{12}$ such that the Poisson structure  
$\{L_1\,,\,L_2\}=[r''_{12}\,,\,L_1]-[r''_{21}\,,\,L_2]$ with $L$ given in \eqref{eq:LaxNov}, yield a consistent Poisson structure for the $(p_i,q_j)$ variables. For $n>2$, the only solution is given by $r''_{12}= x\,(a''_{12}+b''_{12}-c''_{12}-d''_{12})$ which is identically zero due to the regularity relation \eqref{eq:trace0}. The calculation was done using a symbolic computation software for $n$ running from 3 to 5.
Hence, in the generic case, we conjecture that there is no non-trivial linear structure, at least directly associated to the quadratic one. Again, the existence (and the exact form) of such 
linear $r$-matrix structure for general Novikov peakons remains an open question.

In the particular case $n=2$, there is indeed a solution related to the solution 
\begin{equation}
r''_{12}= \frac12\Big(\,a''_{12}-b''_{12}-c''_{12}+d''_{12}\Big)=\,a''_{12}-c''_{12}
=\begin{pmatrix} 1 &0 &0 &0 \\ 2& 0& -1 &0 \\ 0 &1 &0 &0 \\ 2 &0 &-2 &1\end{pmatrix}\,.
\end{equation}
The $r$-matrix obeys the modified Yang--Baxter relation
\begin{equation}
[ r''_{12},r''_{13} ] + [ r''_{12},r''_{23} ] + [ r''_{32},r''_{13} ] = \Omega_{123} - \Omega_{123}^{t_1t_2t_3}
\end{equation}
and leads to PB of the form
\begin{equation}
\begin{aligned}
\{\bar p_1\,,\,\bar p_2\}&=-4\,\fs_{12}\sqrt{\bar p_1\bar p_2}\,e^{-\frac12|\bar q_1-\bar q_2|} \,,\\
\{\bar q_1-\bar q_2\,,\,\bar p_1\}&=4\,\sqrt{\frac{\bar p_1}{\bar p_2}}\,e^{-\frac12|\bar q_1-\bar q_2|}+4 \,,\\
\{\bar q_1-\bar q_2\,,\,\bar p_2\}&=-4\,\sqrt{\frac{\bar p_2}{\bar p_1}}\,e^{-\frac12|\bar q_1-\bar q_2|}-4 \,.
\end{aligned}
\end{equation}
Indeed since the combination $\bar q_1+\bar q_2$ does not appear in the expression of $L$, one can realize a consistent associative Poisson bracket by setting $\{\bar q_1+\bar q_2\,,\,X\}=0$ for all $X$.

\subsection{Dual presentation for Novikov peakons}
Let us remark that the form of the Novikov Lax matrix $L^N=T\,L^{CH}$ suggests a dual presentation for the Novikov peakons.
Indeed, one can introduce the Lax matrix $\wt L^N=L^{CH}\,T$, which takes explicitly the form
\begin{equation}
\wt L_{ij} = \sum_{k=1}^n \sqrt{\bar{p}_k\,\bar{p}_i}\, e^{-\half|\bar{q}_i-\bar{q}_k| }\,T_{kj} \,.
\end{equation}
In that case, the PB \eqref{eq:PBNov} have still a quadratic structure of the form \eqref{eq:L1L2Nov}, but with now
\begin{equation}
\begin{aligned}
& \wt a''_{12} =  4\,a_{12}, \qquad && \wt d''_{12} = 4\, T^{-1}_1\,T^{-1}_2\, a_{12}\, T_1\,T_2, \\
& \wt b''_{12} = T_1^{-1} \, \big( -4a_{12}^{t_2}- \cQ_{12}\big)\, T_1, \qquad &&\wt c''_{12} = \wt b''_{21}.
\end{aligned}
\end{equation}
It is easy to see that these matrices still obey precisely the same Yang--Baxter relations \eqref{eq:ybe-Novi}.

The Hamiltonians one constructs using $\wt L^N$ are exactly the same as for $L^N$. Thus, we  get a dual presentation of exactly the same model and the same Hamiltonians. This property extends to the calculation presented in the next section.

\section{Non-trivial boundary terms for peakons \label{sect:twist}}
It is known that other sets of Hamiltonians can be defined for each solution $\gamma$ of the dual classical reflection equation
\begin{equation}\label{eq:class-RE}
 \gamma_1 \gamma_2 a_{12} - d_{12} \gamma_1 \gamma_2  + \gamma_2 b_{12} \gamma_1 - \gamma_1 c_{12} \gamma_2=0\,.
\end{equation}
The Hamiltonians take the form $\tr\big((\gamma L)^k\big)$.
The solution $\gamma=\II_n$ exists whenever the  condition \eqref{eq:trace0} holds, motivating its designation as \textsl{regularity condition}.
Note that in the Freidel--Maillet approach \cite{MF}, the immediate correspondence between solutions $\gamma$ of the dual equation and  solutions $\gamma^{-1}$ of the direct equation (r.h.s. of \eqref{eq:PBFM} = 0) is used to yield an equivalent form of the commuting Hamiltonians.

We shall now propose classes of invertible solutions to the classical reflection equation
\eqref{eq:class-RE} for each of the three peakon cases.
Let us first emphasize that any solution $\gamma$ of \eqref{eq:class-RE} where $a'_{12}=4a_{12}$, $b'_{12}$, $c_{12}=b'_{21}$, $d'_{12}=4a_{12}$ are the matrices associated with the Degasperis--Procesi peakons, or to the alternative presentation of Camassa--Holm peakons, corresponds to a solution $\gamma'=\gamma T^{-1}$ for the reflection equation associated to the Novikov peakons, since the structure matrices are related as: 
$b_{12}'' = 2T_2\,b'_{12}\,T_2^{-1}$, $c_{12}=b''_{21}$,
 $a''_{12}=2T_1\,T_2\,a'_{12}\,T_1^{-1}\,T_2^{-1}$ and $d''_{12}=2a'_{12}$.
Hence the Degasperis--Procesi case provides solutions for the two other peakon models.

\begin{lemma}\label{lem:diag}
If $\gamma$ is a solution of the Degasperis--Procesi reflection equation \eqref{eq:class-RE}, then for any diagonal matrix $D$, 
the matrix $D\gamma D$ is also a solution of the reflection equation.
\end{lemma}
\proof
The structure matrices $a'_{12}$ and $b'_{12}$ of \eqref{eq:class-RE} for the Degasperis--Procesi peakons obey :
\begin{align}
 a'_{12}\,D_1D_2=D_1D_2\,a'_{12} \qquad \text{and}\qquad
 D_1\,b'_{12}\,D_2=D_2\,b'_{12}\,D_1\,,
\end{align}
where we used also the property \eqref{eq:propQ}.
Now, multiplying \eqref{eq:class-RE} by $D_1D_2$ on the right or / and the left hand sides, and using the above properties of $a'_{12}$ and $b'_{12}$,  leads to the desired result.
\qed
Note that the transformation $\gamma\,\to\,D\gamma D$ is equivalent to a canonical redefinition $p_i\,\to\, d^2_i\,p_i$.
\begin{prop}\label{prop:twist}
For Degasperis--Procesi peakons, and for $n$ arbitrary, we have two fundamental solutions: \\
\null\qquad(i) the unit matrix $\II_n$, \\
\null\qquad(ii) the matrix $T$ introduced in \eqref{eq:defT}.\\
Moreover, when $n$ is even, we have an additional solution  whose explicit form depends on the Weyl chamber we consider for the variables $q_i$. In the first Weyl chamber, where $q_i>q_j\ \Leftrightarrow\ i>j$, it takes the form
\begin{equation}
(iii)\qquad S^{id}=\sum_{i=0}^{\frac{n}2-1} \big(e_{2i+1,2i+2} -e_{2i+2,2i+1}\big)=\II_{n/2}\otimes\, \begin{pmatrix} 0 & 1 \\ -1 & 0 \end{pmatrix}. 
\end{equation}
In any other Weyl chamber defined by a permutation $\sigma$ such that $q_i>q_j\ \Leftrightarrow\ \sigma(i)>\sigma(j)$,
the solution $S^\sigma$ takes the form
\begin{equation}
S^{\sigma}=\sum_{i=0}^{\frac{n}2-1} \big(e_{\sigma(2i+1),\sigma(2i+2)} -e_{\sigma(2i+2),\sigma(2i+1)}\big). 
\end{equation}

Using the lemma \ref{lem:diag}, it leads to 2 (resp. 3) classes of solutions for $n$ odd (resp. even).

All these solutions are also valid for the alternative presentation of Camassa--Holm, and (once multiplied on the left by $T$) for the Novikov model.
The solutions (i) and (iii) are also valid for the original Camassa--Holm model.
\end{prop}
\proof
$(i)$ The unit matrix is trivially a solution since \eqref{eq:class-RE} is then the regularity condition. \\ 
$(ii)$ The reflection equation for $T$ projected on a generic element $e_{ij} \otimes e_{kl}$ contains explicitly the indices $i,j,k,l$, 
and possibly two summation indices corresponding to the products by $\gamma_1$ and $\gamma_2$. Since the entries of the matrices depend on the indices 
only through the sign function $\sgn(r-s)$, it is sufficient to check the relations for small values of $n$. We  verified them through a symbolic calculation software for $n$ running from 2 to 8. \\
$(iii)$ Similarly, the reflection equation for $S$ needs to be checked for small values of $n$. We  verified it through a symbolic calculation software for $n$ running from 2 to 8.  
\qed
\begin{rmk}\label{lem:transp-inv}
A solution $\gamma$  to the reflection equation also leads to solutions of the form $\gamma^t$ and $\gamma^{-1}$ for Camassa--Holm peakons. However, the classes of solutions  they lead to, falls in the ones already presented in proposition \ref{prop:twist}.
\end{rmk}

Note that dressing $T$ by the diagonal matrix $D=\diag((-1)^i)$ yields $T^{-1}$, which is also a solution to the Degasperis--Procesi reflection equation. Moreover, it proves that the unit matrix is also a solution of the reflection equation \eqref{eq:class-RE} in the Novikov case, proving the property mentioned in the previous section that regularity condition is fulfilled by the Novikov $r$-matrix structure.

\section{Hamiltonians}
Now that the quadratic $r$-matrix structure have been defined, we are in position to compute higher Hamiltonians for each of the three classes of peakons. 
These Hamiltonians will be PB-commuting, with the Poisson brackets \eqref{eq:PBCH}, \eqref{eq:PBDGP} or \eqref{eq:PBNov}, 
depending on the peakon model that is to say the Lax matrices \eqref{eq:LCH}, \eqref{eq:LaxGP} or \eqref{eq:LaxNov}.
We provide also some cases of Hamiltonians with non-trivial boundary terms.
\subsection{Camassa--Holm Hamiltonians}
In addition to the peakon Hamiltonian 
$H_{CH}=\tr L=\sum_i p_i$, we get for instance
\begin{equation}\label{eq:Hn-CH}
\begin{aligned}
H^{(1)}_{CH}&=\tr L=\sum_i p_i\,,\\
H^{(2)}_{CH}&=\tr L^2=\sum_{i,j} p_ip_j \,e^{-|q_i-q_j|} \,,\\
H^{(3)}_{CH}&= \tr L^3=\sum_{i,j,k} p_ip_jp_k \,e^{-\frac12|q_i-q_j|}\,e^{-\frac12|q_j-q_k|}\,e^{-\frac12|q_k-q_i|}\,.
\end{aligned}
\end{equation}
We recognize in $H^{(1)}_{CH}$ and $H^{(2)}_{CH}$ the usual Camassa--Holm Hamiltonians, as computed e.g. in \cite{dGHH}.
\paragraph{Diagonal boundary term.} If one chooses $\gamma=D$ as a diagonal solution to the reflection equation, we get another series of PB-commuting Hamiltonians:
\begin{equation}
\begin{aligned}
\tr(DL)&=\sum_i d_i\,p_i\,,\\
\tr \big((DL)^2\big)&=\sum_{i} d_i^2\,p_i^2 +2\sum_{i<j} d_i\,d_j\,p_ip_j \,e^{-|q_i-q_j|} \,,\\ 
\tr \big((DL)^3\big)&=\sum_{i,j,k} d_i\,d_j\,d_k\,p_ip_jp_k \,e^{-\frac12|q_i-q_j|}\,e^{-\frac12|q_j-q_k|}\,e^{-\frac12|q_k-q_i|}.
\end{aligned}
\end{equation}
One gets a "deformed" version of the Camassa--Holm Hamiltonians, with deformation parameters $d_i$.
\paragraph{$T$-boundary term.} Choosing now $\gamma=DTD$ as a  solution to the reflection equation, we get:
\begin{equation}
\begin{aligned}
\tr(\gamma\,L)&=\sum_{i} d_i^2\,p_i +2\sum_{i<j} d_id_j\,\sqrt{p_ip_j} \,e^{-\frac12|q_i-q_j|} \,,\\
\tr \big((\gamma\,L)^2\big)&=\sum_{i} d_i^4\,p_i^2+3 \sum_{i\neq j} d^2_i\,d^2_j\,p_ip_j \,e^{-|q_i-q_j|} 
+4 \sum_{i\neq j} d^3_i\,d_j\,\sqrt{p_i^3p_j} \,e^{-\frac12|q_i-q_j|} \\
&+6\sum_{\atopn{i, j,k}{\text{all }\neq}} d_i^2d_jd_k\,p_i\sqrt{p_jp_k} \,e^{-\frac12|q_i-q_j|}\,e^{-\frac12|q_i-q_k|}\\
&+\sum_{\atopn{i, j,k,l}{\text{all }\neq}} (1+\fs_{jk}\fs_{l i})\,d_id_jd_kd_l\,\sqrt{p_ip_jp_kp_l} \,e^{-\frac12|q_i-q_j|}\,e^{-\frac12|q_k-q_l|} \,.\\ 
\end{aligned}
\end{equation}
Note that since the alternative presentation of Camassa--Holm peakons describes the same Poisson structure, the above Hamiltonians are also valid when using the presentation of section \ref{sect2}.

\paragraph{$S$-boundary term for $n$ even.} Since for Camassa--Holm peakons, the Lax matrix  $L$ is symmetric while $S^\sigma$ is antisymmetric, we get in any Weyl chamber
\begin{equation}
\tr\big((S^\sigma L)^{2m+1}\big)=0\,,\forall m\,.
\end{equation}
As an example of non-vanishing Hamiltonian, we have for $\gamma=DS^\sigma D$ (in the Weyl chamber defined by $\sigma$):
\begin{equation}
\tr\big((\gamma L)^{2}\big)=2\sum_{\ell=0}^{n/2-1} d_{\ell'}d_{\ell'+1}\,p_{\sigma(\ell')}\,p_{\sigma(\ell'+1)}\,
\Big( e^{-|q_{\sigma(\ell'+1)}-q_{\sigma(\ell')}|}- 1\Big)\,,
\end{equation}
where $\ell'=2\ell+1$.

\subsection{Degasperis--Procesi Hamiltonians}
\paragraph{Diagonal boundary term.}  Considering immediately the case with diagonal matrix $\gamma$, we have 
\begin{equation}\label{eq:Hn-DP}
\begin{aligned}
\tr (\gamma L)&=\sum_i d_i\,p_i\,,\\
\tr\big( (\gamma L)^2\big)&=\sum_{i} d_i^2\,p_i^2+\sum_{i<j} d_i\,d_j\,p_ip_j \,\Big(2-e^{-|q_i-q_j|}\Big)e^{-|q_i-q_j|} \,,\\
 \tr\big( (\gamma L)^3\big)&= \sum_{i,j,k} d_id_jd_k\,p_ip_jp_k \, \Big( -3\,e^{-|q_i-q_k|-|q_j-q_k|}  
   + 4 \,e^{- \frac12 (|q_i-q_k|+|q_j-q_k|+|q_j-q_i|)} \Big)\,.
\end{aligned}
\end{equation}
The "usual" Hamiltonians $\tr L^m$ are recovered by setting $d_i=1$, $\forall i$. 
\paragraph{$T$-boundary term.} 
In any Weyl chamber, we get
\begin{equation}\label{eq:Hn-DP}
\begin{aligned}
\tr (\gamma T \gamma L)&=\sum_{i} d_i\,p_i+\sum_{i\neq j}  d_i\,d_j\,\sqrt{p_i\,p_j}\,e^{-|q_i-q_j|}\,,  \\
\tr \big((\gamma T \gamma L)^2\big)&=\Big(\sum_{i, j}  d_i\,d_j\,\sqrt{p_i\,p_j}\Big)^2
+\sum_{i\neq j}  d^2_i\,d^2_j\,p_i\,p_j\,\big(1-e^{-|q_i-q_j|}\big)^2 \\
&-4\sum_{i\neq j} d_i^2\,d_j\,p_i\,\sqrt{p_j}\big(1-e^{-|q_i-q_j|}\big)\Big(\sum_k d_k\sqrt{p_k}\Big)\\
&-8\sum_{q_j<q_i<q_k} d_id_jd_k\,\sqrt{p_ip_jp_k}\big(1-e^{-|q_j-q_k|}\big)\Big(\sum_l d_l\sqrt{p_l}\Big)\\
&+2\sum_{\atopn{i,j,k}{\text{all }\neq}} d_i^2\,d_jd_k\,p_i\sqrt{p_jp_k}\, \big(1-e^{-|q_i-q_j|}\big)\big(1- e^{- |q_i-q_k|} \big)\\
&+8\sum_{q_i<q_j<q_k<q_l}d_id_jd_kd_l\,\sqrt{p_ip_jp_kp_l}\, \big(1-e^{-|q_i-q_l|}\big)\big(1- e^{- |q_j-q_k|} \big)
\, .
\end{aligned}
\end{equation}

\paragraph{$S$-boundary term for $n$ even.} 
In the Weyl chamber characterized by $\sigma$, wet get:
\begin{equation}
\begin{aligned}
\tr(S^\sigma L)&=2\sum_{\ell=0}^{\frac n2-1}\,d_{\ell+1}\,\sqrt{p_{\sigma(\ell')}\,p_{\sigma(\ell'+1)}}\Big(1-e^{-|q_{\sigma(\ell')}-q_{\sigma(\ell'+1)}|}\Big)\,,\\
\tr\big((S^\sigma L)^2\big)&=2\sum_{\ell=0}^{\frac n2-1}\,d_{\ell+1}^2\,p_{\sigma(\ell')}\,p_{\sigma(\ell'+1)}\Big(1-e^{-|q_{\sigma(\ell')}-q_{\sigma(\ell'+1)}|}\Big)^2\\
&+
2\sum_{\ell=0}^{\frac n2-1}\sum_{\atopn{j=0}{j\neq\ell}}^{\frac n2-1}\,d_{\ell+1}d_{j+1}\,\sqrt{p_{\sigma(\ell')}\,p_{\sigma(\ell'+1)}p_{\sigma(j')}\,p_{\sigma(j'+1)}}\\
&\qquad\times\Big(2e^{-|q_{\sigma(\ell'+1)}-q_{\sigma(j')}|} -e^{-|q_{\sigma(\ell'+1)}-q_{\sigma(j'+1)}|}-e^{-|q_{\sigma(\ell')}-q_{\sigma(j')}|}\Big)\end{aligned}
\end{equation}
where we noted $\ell'=2\ell+1$ and $\jcur'=2j+1$ to have more compact expressions.

\subsection{Novikov Hamiltonians}
We recall that the Novikov and Camassa--Holm Lax matrices are related by $L^{Nov}=T\,L^{CH}$ and that the solutions of the dual reflection equation \eqref{eq:class-RE} are related dually: $\gamma^{Nov}=\gamma^{CH}\,T^{-1}$.
The basic combination $\gamma L$ entering the Hamiltonians for the Novikov peakons therefore yields the same object as for the Camassa--Holm case. Hence Camassa--Holm and Novikov peakons share identical forms of commuting Hamiltonians, a consistent consequence of their sharing the same Poisson structure \eqref{eq:PBNov2} after renormalization \eqref{eq:RenNov}.

\section{Conclusion}
Having established the quadratic Poisson structures for three integrable peakon models, in every case based on the Toda molecule $r$-matrix and its partial transposition, many issues remain open or have arisen in the course of our approach.

Amongst them we should mention the problem of finding compatible linear Poisson structures and their underlying $r$-matrix structures. Only solved for the Camassa Holm peakons this question is particularly subtle in the case of Degasperis Procesi peakons as follows from the discussion in Section $3.3$.

Another question left for further studies here is the extension of integrability properties beyond the exact peakon dynamics, on the lines of the work in \cite{RB} regarding Camassa Holm peakons and (not surprisingly) based on the linear (and canonical) Poisson structure, yet unidentified in other cases. The difficulty is to disentangle , whenever only a quadratic structure is available, the peakon potential in the Lax matrix from the dynamical weight function in the $n$-body Poisson brackets (the so called $G$ function in \cite{dGHH}).

Excitingly, the quadratic structures we have found open the path to a quantized version of peakon models, in the form of $ABCD$ algebras, following  the lines developed in \cite{MF}. We hope to come back on this point in a future work.

Finally in this same extended integrable peakons the still open problem of full understanding of the unavoidably dynamical $r$-matrix structure remains a challenge.



\begin{thebibliography}{99}

\bibitem{BV}
O. Babelon, C. Viallet,
\textsl{Hamiltonian structures of Lax equations,}
Phys. Lett. B \textbf{237}, 411 (1989), \doi{10.1016/0370-2693(90)91198-K}.

\bibitem{STS}
M.A. Semenov-Tyan-Shanskii,
\textsl{What is a classical $r$-matrix?,}
Funct. Anal. Appl. \textbf{17}, 259 (1983), \doi{10.1007/BF01076717}.

\bibitem{Skl1982}
E.K. Sklyanin,
\textsl{Some algebraic structures connected with the Yang--Baxter equation,}
Funct. Anal. Appl. \textbf{16}, 263 (1982), \doi{10.1007/BF01077848}.

\bibitem{Skl1983}
E.K. Sklyanin,
\textsl{Some algebraic structures connected with the Yang—Baxter equation. Representations of quantum algebras,}
 Funct. Anal. Appl. \textbf{17}, 273 (1983), \doi{10.1007/BF01076718}.

\bibitem{MF}
L. Freidel, J.-M. Maillet,
\textsl{Quadratic algebras and integrable systems,}
Phys. Lett. B \textbf{262}, 278 (1991), \doi{10.1016/0370-2693(91)91566-E}.

\bibitem{Skl1979}
E.K. Sklyanin,
\textsl{On the complete integrability of the Landau--Lifshitz equation,}
preprint LOMI E-3-1979, Leningrad, LOMI (1980) [not published].

\bibitem{AR2016}
S.C. Anco, E. Recio,
\textsl{A general family of multipeakon equations,}
J. Phys. A: Math. Theor. \textbf{52}, 125203 (2019), \doi{10.1088/1751-8121/ab03dd}, \texttt{arXiv:1609.14354 [math-ph]}.

\bibitem{CH}
R. Camassa, D.D. Holm,
\textsl{An integrable shallow water equation with peaked solitons,}
Phys. Rev. Lett. \textbf{71}, 1661 (1993), \doi{10.1103/PhysRevLett.71.1661}, \texttt{arXiv:patt-sol/9305002}.

\bibitem{CHH}
R. Camassa, D.D. Holm, J.M. Hyman,
\textsl{A new integrable shallow wave equation,}
Adv. Appl. Mech. \textbf{31}, 1 (1994), \doi{10.1016/S0065-2156(08)70254-0}.

\bibitem{dGHH}
A. Degasperis, D.D. Holm, A.N.W. Hone,
\textsl{Integrable and non-integrable equations with peakons,}
Proceedings of Nonlinear Physics II: Theory and Experiment, Gallipoli 2002, pp. 37--43, ed. World Scientific, \doi{10.1142/9789812704467$\_$0005}, \texttt{arXiv:nlin/0209008 [nlin.SI]}.

\bibitem{RB}
O. Ragnisco, M. Bruschi,
\textsl{Peakons, $r$-matrix and the Toda lattice,}
Physica A: Stat. Mech. Appl. \textbf{228}, 150 (1996), \doi{10.1016/0378-4371(95)00438-6}, \texttt{arXiv:solv-int/9509012}.

\bibitem{DGP}
A. Degasperis, M. Procesi,
\textsl{Asymptotic integrability,}
in: \textsl{Symmetry and Perturbation Theory} (A. Degasperis and G. Gaeta, eds.), World Scientific, Singapore (1999), pp. 23--37.

\bibitem{dGPro3}
A. Degasperis, D.D. Holm, A.N.W. Hone,
\textsl{A new integrable equation with peakon solutions,}
Theor. Math. Phys. \textbf{133}, 1463 (2002), \doi{10.1023/A:1021186408422}, \texttt{arXiv:nlin/0205023 [nlin.SI]}.

\bibitem{HH}
D.D. Holm, A.N.W. Hone,
\textsl{A class of equations with peakons and pulson solutions (with an Appendix by Harry Braden and John Byatt-Smith),}
J. Nonlin. Math. Phys. \textbf{12}, 380 (2005), \doi{10.2991/jnmp.2005.12.s1.31}, \texttt{arXiv:nlin/0412029 [nlin.SI]}.

\bibitem{Novi1}
V.S. Novikov,
\textsl{Generalizations of the Camassa--Holm equation,}
J. Phys. A: Math. Theor. \textbf{42}, 342002 (2009), , \doi{10.1088/1751-8113/42/34/342002}, \texttt{arXiv:0905.2219 [nlin.SI]}.

\bibitem{HW}
A.N.W. Hone, J.P. Wang,
\textsl{Integrable peakons with cubic linearity,}
J. Phys. A: Math. Theor. \textbf{41}, 372002 (2008), , \doi{10.1088/1751-8113/41/37/372002}, \texttt{arXiv:0805.4310 [nlin.SI]}.

\bibitem{HLS}
A.N.W. Hone, H. Lundmark, J. Szmigielski, 
\textsl{Explicit multipeakon solutions of Novikov's cubically nonlinear integrable Camassa--Holm type equation,}
Dyn. PDE \textbf{6}, 253 (2009), \doi{10.4310/DPDE.2009.v6.n3.a3}, \texttt{arXiv:0903.3663}.

\bibitem{Fuchs}
B. Fuchssteiner,
\textsl{Some tricks from the symmetry-toolbox for nonlinear equations: Generalizations of the Camassa--Holm equation,}
Physica D: Nonlinear Phenomena \textbf{95}, 229 (1996), \doi{10.1016/0167-2789(96)00048-6}.

\bibitem{Fokas}
A.S. Fokas,
\textsl{The Korteweg--de Vries equation and beyond,}
Acta Appl. Math. \textbf{39}, 295 (1995), \doi{10.1007/BF00994638}.

\bibitem{AK}
S.C. Anco, D. Kraus,
\textsl{Hamiltonian structure of peakons as weak solutions for the modified Camassa--Holm equation,}
Discrete and Continuous Dynamical Systems (Series A) \textbf{38}, 4449 (2018), \doi{10.3934/dcds.2018194}, \texttt{arXiv:1708.02520 [nlin.SI]}.

\bibitem{OT}
D. Olive, N. Turok,
\textsl{Algebraic structure of Toda systems,}
Nucl. Phys. B \textbf{220}, 491 (1983), \doi{10.1016/0550-3213(83)90504-7}.

\bibitem{FO}
L. Ferreira, D. Olive,
\textsl{Non-compact symmetric spaces and the Toda molecule equations,}
Commun. Math. Phys. \textbf{99}, 365 (1985), \doi{10.1007/BF01240353}.

\bibitem{Magri}
F. Magri,
\textsl{A simple model for the integrable Hamiltonian equation,}
J. Math. Phys. \textbf{19}, 1156 (1978), \doi{10.1063/1.523777}.

\bibitem{OR}
W. Oevil, O. Ragnisco,
\textsl{$R$-matrices and higher Poisson brackets for integrable systems,}
Physica A: Stat. Mech. Appl. \textbf{161}, 181 (1989), \doi{10.1016/0378-4371(89)90398-1}.

\bibitem{CF}
F. Calogero and J.‐P. Françoise,
\textsl{A completely integrable Hamiltonian system,}
J. Math. Phys. \textbf{37}, 2863 (1996), \doi{10.1063/1.531536}.

\bibitem{ABT}
J. Avan, O. Babelon, M. Talon,
\textsl{Construction of the classical $R$-matrices for the Toda and Calogero models,}
Algebra i Analiz \textbf{6}, 67 (1994), \texttt{arXiv:hep-th/9306102}.

\bibitem{AR}
J. Avan, E. Ragoucy,
\textsl{Rational Calogero-Moser Model: Explicit Form and r-Matrix of the Second Poisson Structure,}
SIGMA \textbf{8}, 079 (2012), \doi{10.3842/SIGMA.2012.079}, \texttt{arXiv:1207.5368 [math-ph]}.

\bibitem{MN}
A.V. Mikhailkov, V.S. Novikov,
\textsl{Perturbative symmetry approach,}
J. Phys. A: Math. Gen. \textbf{35}, 4775 (2002), \doi{10.1088/0305-4470/35/22/309}, \texttt{arXiv:nlin/0203055}.







\end{thebibliography}
\end{document}